\documentclass[preprint,showpacs,preprintnumbers,amsmath,amssymb,prb]{revtex4}
\usepackage{amsmath}
\usepackage{graphicx}% Include figure files
\usepackage{dcolumn}% Align table columns on decimal point
\usepackage{bm}% bold math
\usepackage{graphics}
\usepackage{epsfig}
\usepackage{subfigure}
\usepackage{color}

\newcommand\eq{\begin{equation}}
\newcommand\be{\begin{equation}}
\newcommand\eeq{\end{equation}}
\newcommand\ee{\end{equation}}
\newcommand\ar{\begin{eqnarray}}
\newcommand\ear{\end{eqnarray}}

\newcommand{\BE}{{\bf E}}
\newcommand{\BH}{{\bf H}}

\newcommand{\ii}{{\rm i}}

\newcommand{\xhat}{{\bf \hat{x}}}

\renewcommand{\Im}{\mbox{Im}}
\renewcommand{\Re}{\mbox{Re}}

\renewcommand{\exp}{\mbox{exp}}
\begin{document}
\newcommand{\nm}{\mbox{ nm}}
\newcommand{\micron}{\mbox{ $\mu$m}}
\newcommand{\fm}{\mbox{ fm}}
\newcommand{\Hz}{\mbox{ Hz}}
\newcommand{\watt}{\mbox{ W}}
\newcommand{\m}{\mbox{ m}}
\newcommand{\cm}{\mbox{ cm}}
\newcommand{\km}{\mbox{ km}}
\newcommand{\seconds}{\mbox{ s}}
\newcommand{\minutes}{\mbox{ min}}
\newcommand{\volts}{\mbox{ V}}
\newcommand{\MeV}{\mbox{ MeV}}
\newcommand{\eV}{\mbox{ eV}}
\newcommand{\meV}{\mbox{ meV}}
\newcommand{\rad}{\mbox{ radians}}
\newcommand{\Kelvin}{\mbox{ K}}
\newcommand{\Siemen}{\mbox{ S}}
\newcommand{\angstrom}{\mbox{ $\AA$}}

\title{Synthesis of highly confined surface plasmon modes with doped graphene sheets in the mid-infrared and terahertz frequencies}

\author{Choon How Gan, Hong Son Chu, and Er Ping Li}

\affiliation{Electronics and Photonics Department,\\
A*STAR Institute of High Performance Computing, 138632, Singapore}

\begin{abstract}
%We report analytical and numerical investigations demonstrating that monolayers and parallel pairs of doped graphene support ultra-confined surface plasmon modes in the mid-infrared and terahertz regime. 
We investigate through analytic calculations the surface plasmon dispersion relation for monolayer graphene sheets and a separated parallel pair of graphene monolayers. 
An approximate form for the dispersion relation for the monolayer case was derived, which was shown to be highly accurate and offers intuition to the properties of the supported plasmon mode.
For parallel graphene pairs separated by small gaps, the dispersion relation of the surface plasmon splits into two branches, one with a symmetric and the other with an antisymmetric magnetic field across the gap. 
For the symmetric (magnetic field) branch, the confinement may be improved at reduced absorption loss over a wide spectrum, unlike conventional SP modes supported on metallic surfaces that are subjected to the trade-off between loss and confinement. This symmetric mode becomes strongly suppressed for very small separations however. On the other hand, its antisymmetric counterpart exhibits reduced absorption loss for very small separations or long wavelengths, serving as a complement to the symmetric branch. 
Our results suggest that graphene plasmon structures could be promising for waveguiding and sensing applications in the mid-infrared and terahertz frequencies.
\end{abstract}
\maketitle

\section{introduction}
Surface plasmons~\cite{raether} (SPs) are charge density waves propagating at the interface between a conductor and a dielectric medium. SPs are transverse-magnetic (TM) polarized with modal fields that peak at the interface and decay exponentially away into both media. Noble metals such as gold and silver have traditionally been the material of choice to support SPs, providing for lossy but well-confined plasmon fields at visible frequencies. To achieve SP resonance in the near-infrared regime and beyond, metallic films corrugated  with nanoholes~\cite{spoof} or nanorods,~\cite{kabashin} and multi-layered metmaterial structures~\cite{elser,ganol} have been proposed. More recently graphene, a single layer of carbon atoms gathered in honeycomb lattice, has also been 
considered for supporting surface plasmon modes~\cite{ranaieee,jablanprb80,engheta,mceuen,abajoacsn12} at infrared and terahertz frequencies. The transport properties of graphene can be readily tuned by application of a gate voltage~\cite{novo666,schedin}, opening the possibilities to various plasmonic devices~\cite{engheta,ferrarinphot,koppens,junatnano,avouris}. Moreover, several techniques have been proposed to experimentally realize the excitation of plasmons on graphene~\cite{efimovprb84, basovnl11}.

The SP dispersion of isolated monolayer graphene ($\sim 0.34 \nm$ thick) and monolayer graphene deposited on dielectric substrates have been studied by several authors.~\cite{hansonjap103,wunschnjp,sarmaprb75} The plasmon dispersion for parallel graphene pairs,~\cite{sarmaprb75,sarmaprb80} multilayer graphene stack,~\cite{falkovskyprb76} and intercalated graphite~\cite{shungprb34} have also been theoretically investigated. 
For a parallel pair, the geometry approaches the bilayer graphene when their separation is small (few angstroms), and effects of interlayer hopping could become important~\cite{borghiprb80,sarmaprb82}.

The transport properties of graphene can be modeled with the linear response theory(Kubo formula).~\cite{hansonjap103,falkovskyprb76,tobiasprb78,gusyninprb75} 
In the nonlocal random phase approximation (RPA), analytical expressions that are valid for arbitrary frequency $\omega$ and plasmon propagation wave vector $\vec{\beta}$ may be derived in the limit of zero temperature.~\cite{ranaieee,wunschnjp,sarmaprb75,borghiprb80,sarmaprb82} 
Alternatively in the local limit where spatial dispersion effects are negligible,~\cite{falkovskyprb76} i.e., $\omega >> \tau^{-1}$ and $\omega \gtrsim \beta v_F$, analytical expressions for the optical conductivity ($\sigma$) of graphene can be obtained from the Kubo formula for finite temperatures $T$. 
Here $\tau$ is a phenomenological electron relaxation time, $v_F \approx 10^6~{\rm m/s}$ is the Fermi velocity, 
and $\beta$ = $|\vec{\beta}|$. 
As dynamic control of the SP dispersion holds potential promise for the design of various active devices such as optical switches and modulators, it is important to study the characteristics of the SP modes such as their confinement and spatial distribution of dominant field components, so as to gain further understanding of their physical properties.   
In this work, further to analyzing the SP dispersion, we describe within the framework of electromagnetic waveguide theory, how the properties of the SP modes supported 
on monolayer graphene and parallel graphene sheets may be tuned over a broad range of frequencies for potential waveguide applications. The characteristics of the SP modes are analyzed through their associated electromagnetic mode profiles.
In addition, we introduce a figure-of-merit (FOM), which in our opinion represents an optimal measure that takes into account the compromise between confinement and propagation loss of the supported SP modes  to quantify the performance of the waveguide.

To describe the transport properties of graphene in our study, we calculate the local conductivity of graphene from the Kubo formula. Let us define the effective index of the supported plasmon mode as $n = \beta/k_0$, where $k_0 = 2\pi/\lambda$ and $\lambda$ is the free-space wavelength. In the local limit ($\omega >> \tau^{-1}$), the condition $\beta v_F \lesssim \omega$ translates to $n \lesssim c/v_F \approx 300$ ($c$ being the speed of light in vacuum), and the expression for $\sigma$ is~\cite{gusyninprb75,falkovskyprb76}
\begin{align}\label{kubo}
\sigma & = \frac{e^2(\omega + \ii \tau^{-1})}{\ii \pi \hbar^2} \Bigg[ \frac{1}{(\omega + \ii \tau^{-1})^2} \int_0^\infty \varepsilon \left( \frac{\partial F(\varepsilon)}{\partial \varepsilon } - \frac{\partial F(-\varepsilon)}{\partial \varepsilon }\right) d\varepsilon \nonumber \\
& \qquad \qquad \qquad \,\, \,\,- \int_0^\infty  \frac{F(-\varepsilon) - F(\varepsilon)}{(\omega + \ii \tau^{-1})^2 - 4(\varepsilon /\hbar)^2 } \,\,d\varepsilon \Bigg]  \nonumber \\
& = \sigma^{\rm intra}  + \sigma^{\rm inter} \,,
\end{align}
where $F(\varepsilon) = (1 + {\rm exp}[(\varepsilon -\mu_c)/K_BT])^{-1}$ is the Fermi-Dirac distribution with $\mu_c$ as the chemical potential.
In Eq.~\eqref{kubo}, the first term corresponds to contributions from intraband electron-photon scattering and the second term arises from contributions due to direct interband electron transitions. Integration of the first term leads to~\cite{falkovskyprb76}
\begin{subequations}\label{sigintra}
\begin{align}
\sigma^{\rm intra} & = \frac{2\ii e^2 K_BT}{\pi \hbar^2 (\omega + \ii \tau^{-1})} {\rm ln} [2 \cosh(\frac{\mu_c}{2K_BT})]\,, \label{sigintraa} \\
& = \frac{\ii e^2 \mu_c}{\pi \hbar^2 (\omega + \ii \tau^{-1})} \,,\quad (\mu_c >> K_BT)\,, \label{sigintrab}
\end{align}
\end{subequations} 
For the interband term, we obtain, after evaluating the numerator of the integrand 
\begin{subequations}\label{siginter1}
\begin{align}
\sigma^{\rm inter} & = \frac{e^2 \ii (\omega + \ii \tau^{-1})}{4\pi K_BT} \int_0^\infty  \frac{G(\xi)}{\hbar^2(\omega + \ii \tau^{-1})^2/(2K_BT)^2 - \xi ^2 } \,\,d\xi \,, \label{siginter1a} \\
& = \frac{e^2}{4\hbar} \left[ 1 + \frac{\ii}{\pi} {\rm ln} \frac{\hbar(\omega + \ii \tau^{-1}) - 2\mu_c}{\hbar(\omega + \ii \tau^{-1}) + 2\mu_c} \right]\,,\quad (\mu_c >> K_BT)\,, \label{siginter1b}
\end{align}
\end{subequations}
where $G(\xi) = \sinh \xi/[\cosh(\mu_c/K_BT) + \cosh \xi]$, and $\xi = \varepsilon /K_BT$. 
For finite values of $\tau$, the integrand in Eq.~\eqref{siginter1a} has no poles along the real $\xi-$axis, and is suitable for numerical integration. 
For $\mu_c >> K_BT$, Eq.~\eqref{sigintrab} shows that $\sigma^{\rm intra}$ is directly proportional 
to the chemical potential $\mu_c$, while Eq.~\eqref{siginter1b} shows that $\sigma^{\rm inter}$ diverges logarithmically for $\hbar \omega \approx 2\mu_c$. 
Furthermore, $\Re(\sigma^{\rm inter})$ jumps by an amount of $\sigma_0$ at the interband threshold $\hbar \omega^{\rm inter} = 2\mu_c$, evident of strong absorption loss for $\omega > \omega^{\rm inter}$. 
At high frequencies, $\omega >> (\mu_c, K_BT),$ the graphene conductivity $\sigma$ approaches 
a constant $\sigma_0 \approx e^2/4\hbar$, which is close to the measured minimum conductivity value of $\frac{e^2}{2\pi \hbar}$~\cite{geimnat438,sarmapnas}. 

For all following calculations, it is taken that $T = 300 \Kelvin$ (room temperature, $K_BT \sim 26 \meV$), and that $\mu_c >> K_BT$. In this limit, Eq.~\eqref{sigintrab} and Eq.~\eqref{siginter1b} for the optical conductivity of graphene holds, and the relationship between the chemical potential $\mu_c$ and the carrier density $n_c$ on the monolayer graphene is 
\eq\label{ncdensity}
\mu_c = \sqrt{(\hbar v_F)^2 n_c\pi - (\pi K_BT)^2/3} \approx \hbar v_F \sqrt{n_c \pi} \,.
\eeq
Since carrier density of up to $10^{14} \cm^{-2}$ has been realized in experiments~\cite{yepnas108,efetov}, 
we set the range of chemical potential to $0.2 \eV \leq \mu_c \leq 1.2 \eV$. 
Based on measured values of the carrier mobility~\cite{novo666,efetov,schedin} which ranges from $10^3 - 10^4 \cm^2\volts^{-1}\seconds^{-1}$, the relaxation time  $\tau$ is taken to be in the order of $\sim 10^{-13} \seconds$. In graphene, effects of free carrier absorption~\cite{yucardona} must be taken into account for 
photon energies above the optical phonon energy~\cite{lazzeriprb73} $\hbar \omega^{\rm op} = 0.196 \eV (\lambda^{\rm op} \approx 6.3 \micron)$ 
but below $\hbar \omega^{\rm inter}$. 
For $\omega^{\rm inter} > \omega > \omega^{\rm op}$, conservation of momentum for the free carrier absorption process can be satisfied with the emission of an optical phonon, leading to a significant decrease~\cite{jablanprb80} in the relaxation time $\tau$. A faster relaxation time corresponds to higher absorption losses, i.e., an increase in $\Re(\sigma)$.
The analytical expressions above for $\sigma$ (Eqs.~\eqref{kubo}--\eqref{siginter1}) hold for electron and hole bands which exhibit a linear energy dispersion relation near the zero bandgap points in graphene~\cite{falkovskyusp}. The range of energy for which this linear dispersion relation is valid extends well into the visible spectrum, as shown by a recent experiment~\cite{nairsci320}. To stay within the regime for which our analysis is valid, we limit the calculation of the graphene conductivity $\sigma$ with Eqs.~\eqref{sigintra}~and~\eqref{siginter1} to wavelengths in the range $0.8 \micron \lesssim \lambda \lesssim 50 \micron$ (near-infrared to terahertz regime). 
For long wavelengths, the conductivity is dominated by the intraband term, with the real and imaginary part scaling as $\lambda^2$ and $\lambda$, respectively, in accordance with Eq.~\eqref{sigintrab}. The conductivity decreases with wavelength until it approaches the interband threshold, where $\Re(\sigma) \longrightarrow  \sigma_0$, and $\Im(\sigma)$ becomes negative. 
When $\Im(\sigma)$ is negative, TE (transverse electric)-polarized surface modes~\cite{hansonjap103,zieglerteprl} can be supported on graphene. This mode, whose dispersion lies close to 
the light line, will not be further discussed here. Instead we focus on the TM plasmon modes, which offers the potential for smaller effective mode areas compared to the noble metals, especially in the long wavelength regime where SPs supported on metals are known to be poorly confined. 

The remainder of this paper is organized as follows. In Sec. II, the simplest case where monolayer graphene is sandwiched between two dielectric half-space ($\epsilon_1, \epsilon_2$) is re-visited. 
It is shown that only the antisymmetric mode (labeled as the A mode) is supported in the case when $\epsilon_1$ and $\epsilon_2$ are lossless dielectrics. By antisymmetric, we mean that the tangential magnetic field component exhibits a zero across the monolayer graphene. 
An approximate form for the dispersion relation for the monolayer case is derived, which is shown to be highly 
accurate and offers intuition to the properties of the supported plasmon mode. The case of two layers of monolayer graphene with 
identical chemical potential ($\mu_{1} = \mu_{2}$)  separated by a 
thin dielectric medium ($\epsilon_d$) of thickness $d$ is investigated in Sec. III, where the separation $d$ is exploited as an additional degree of freedom to tune the properties of the SP modes. 
For the fully symmetric geometry ($\mu_{1} = \mu_{2}, \epsilon_1 = \epsilon_2$) and when $\{\epsilon_1, \epsilon_2, \epsilon_d\}$ are purely real, the SP 
dispersion splits into a symmetric AS branch, and an antisymmetric AA branch, where the first A in each label refers to the antisymmetric magnetic field distribution across each monolayer graphene. The second label S (A) refers to the mode with symmetric (antisymmetric) magnetic field across the dielectric medium separating the graphene pair.
The variation of the SP dispersion with wavelength ($\lambda$), chemical potential ($\mu_c$), and the dielectric constant of the separating medium ($\epsilon_d$), will be examined.
The influence on the dispersion for an asymmetric geometry where the chemical potential on each graphene monolayer is different, will also be discussed.
Finally, we offer concluding remarks in Sec. IV.

\section{SP modes on monolayer graphene}
To begin, we analyze the case where monolayer graphene (at $x = 0$) is sandwiched between two dielectric media ($\epsilon_1, \epsilon_2$) as shown in Fig.~1(a).  For a guided wave with wavenumber $\beta$ propagating along the $z$ axis, the tangential component of $H$-field along the $y$ axis can be expressed as
\begin{equation}\label{hfieldtm1}
H = \left \{ \begin{array}{ll}
N_0H_{1} \,\, \exp(\alpha_1 x + \ii \beta z) \, & \textrm {$x < 0$ }\\
N_0H_{2} \,\, \exp(-\alpha_2 x + \ii \beta z) \, & \textrm {$x > 0$ }
\end{array} \right.
\end{equation} 
where $N_0$ is a normalization constant such that the mode has unity Poynting flux at $z = 0$, and 
\eq\label{alpkz}
\alpha_j^2 = \beta^2-\epsilon_j k_0^2\,, \quad j = \{1, 2\}\,. 
\eeq
Here we suppress the time dependence $\exp(-\ii\omega t)$ for notation brevity. The electric field components, evaluated from Maxwell equations are 
\begin{subequations}\label{exezigi}
\begin{align}
E_{x_j} & = \frac{\beta}{\omega\epsilon_j\epsilon_0}H_{j}\,, \label{exigi} \\
E_{z_j} & = \frac{\ii}{\omega\epsilon_j\epsilon_0}\frac{\partial H_{j}}{\partial x}\,. \label{ezigi}
\end{align}
\end{subequations} 
At $x = 0$, the continuity of the tangential electric field ($E_{z_1} = E_{z_2}$) yields
\eq\label{tanecon}
H_{2} = -\frac{\alpha_1\epsilon_2}{\alpha_2\epsilon_1}H_{1},
\eeq
and the boundary condition for the tangential magnetic field, i.e., $\xhat \times (\BH_2 - \BH_1) = \sigma \BE\, (\xhat \equiv$ unit vector along $x$) leads to $H_{2}-H_{1} = \sigma E_z,$ or
\eq\label{tanhcon}
H_{2}  = (1 + \frac{\ii \sigma \alpha_1}{\omega \epsilon_1\epsilon_0}) H_{1}  \,.
\eeq
On substituting Eq.~\eqref{tanecon} into Eq.~\eqref{tanhcon}, we arrive at the dispersion relation
\eq\label{spdispigi}
\frac{\epsilon_1}{\alpha_1} + \frac{\epsilon_2}{\alpha_2} + \frac{\ii \sigma}{\omega\epsilon_0} = 0 \,.
\eeq
The dispersion relation~\eqref{spdispigi} bears a striking resemblance with that for the SPs at a metal-dielectric interface. If either one of the dielectric media is a metal, then it is clear that the SP mode is almost identical to that for the metal-dielectric interface since $|\sigma|$ lies in the range $10^{-4} - 10^{-3} \Siemen$ and $\omega$ is in the range $10^{13} - 10^{15} \seconds^{-1}$. We have verified this fact numerically, which is consistent with experimental observations where carbon-on-gold substrates are used to 
improve the stability in SP resonance detection of DNA arrays~\cite{lockett}. 
Because of the similarity to the classical metal-dielectric interface, the metal-graphene-insulator case is not interesting in the present context and we shall focus only on the insulator-graphene-insulator (I-G-I) structure. To aid in the analysis, we note that the 
intraband conductivity of graphene takes the form similar to the Drude model for free electrons, i.e.,
$\sigma^{\rm intra} = \ii e^2 v_F \sqrt{n_c}/\left[\sqrt{\pi}\hbar(\omega + \ii \tau^{-1})\right]$, and define an effective plasmon frequency
\begin{align}\label{effplasmonfreq}
(\omega_p^{\rm (eff)})^2 & = \frac{\sqrt{n_c} v_F e^2}{\epsilon_0 \hbar \Delta} \nonumber \\
& = \frac{n_c e^2}{\epsilon_0 m_c \Delta}
\end{align}
where $m_c = \hbar \sqrt{n_c \pi} / v_F$ is the effective mass of the Dirac fermions~\cite{geimnat438}, and $\Delta$ is the effective thickness of the monolayer graphene. 
For graphene, $\omega_p^{\rm (eff)}$ serves as an estimate of the plasmon resonance frequency. 
As opposed to metals with parabolic electron dispersion where the plasmon frequency is 
proportional to $n_c^{1/2}$, $\omega_p^{\rm (eff)} \propto n_c^{1/4}$ for graphene.~\cite{junatnano,sarmaprb75} 
Let us note that $\Delta$ in Eq.~\eqref{effplasmonfreq} is not necessarily the physical thickness of the graphene layer, rather it is the effective thickness that would allow us to recover the correct dispersion relation when the monolayer graphene is treated as a thin metal film or screening layer. To estimate $\omega_p^{\rm (eff)}$, we take $\Delta \sim 0.34\nm$, which gives $\lambda_p^{\rm (eff)} = 2\pi c/\omega_p^{\rm (eff)} \sim 300 \nm$ for $0.4 \eV \leq \mu_c \leq 1.2 \eV$.

Far from the plasmon resonance frequency, the loss of the bound modes or equivalently the 
imaginary part of the wavenumber ($\beta = \beta' +\ii \beta''$) is small. 
Consequently, Eq.~\eqref{alpkz} implies that the imaginary part of $\alpha$ is small compared to the real part ($\alpha = \alpha' +\ii \alpha''$), and $\alpha \approx \alpha'$. 
As $\alpha'$ must be positive for the bound modes, Eq.~\eqref{tanecon} implies that the $H$-field must change sign across the monolayer graphene when both the adjacent media are lossless dielectrics, and hence only the antisymmetric mode (labeled A) is supported in the case of the I-G-I. 
Taking $\epsilon_2 = 1$ and $\mu_c = 0.8 \eV$, the real part of the SP mode effective index ($n = n' + \ii n'' = \beta/k_0$)  for $\epsilon_1 = 1$ and 4 is shown in Fig.~1(b). 
Following the derived results of Ref.~[21], we have also calculated the SP dispersion relation with the nonlocal RPA, which shows a good agreement with Eq.~\eqref{spdispigi} (see black-dashed curves in Fig.~1(b)).
Similar to the antisymmetric SP modes supported on thin metal films, it is seen that $n' > \{\sqrt{\epsilon_1}, \sqrt{\epsilon_2}\,\}$. 
As $\lambda \longrightarrow \lambda_p^{\rm (eff)}$, $n'$ increases rapidly.
The electromagnetic field components for $\{\epsilon_1, \epsilon_2\} = \{1, 1\}$ and $\{\epsilon_1, \epsilon_2\} = \{4, 1\}$, are shown in figs.~1(c) and 1(d) 
respectively, taking $\mu_c = 0.8 \eV$ and $\lambda = 8 \micron$ as an example.
It is evident that the $H$-field of the A mode switches sign across the monolayer graphene, and has a maximum at the side of the dielectric with a higher permittivity (if $\epsilon_1 \neq \epsilon_2)$. 
It is worth noting that because $\beta$ is relatively large compared to $k_0$, the magnitude of the normal and tangential $E-$field components are comparable~\cite{raether} according to Eq.~\eqref{exezigi}, i.e., $|E_x| \approx |E_z|$. 
This is in contrast to SP modes supported on metal-air interfaces that have $\beta' \approx k_0$, and have the normal electric field component as the dominant $E-$field component. 
For cases where the wavenumber of the SP supported on monolayer graphene satisfy $\beta' >> k_0$, one should be able to make the approximation $\alpha_j \approx \beta$ (see Eq.~\eqref{alpkz}), which remains accurate provided $\beta'' << \beta'$ and $\alpha'' << \alpha'$. 
Substituting the approximation $\beta \approx \alpha_j$ into Eq.~\eqref{spdispigi} yields a simple and physically intuitive solution
\begin{align}\label{betaapprox}
\beta & \approx  \ii \omega\epsilon_0(\epsilon_1 + \epsilon_2) / \sigma \nonumber \\
& = D (\epsilon_1 + \epsilon_2) (\omega^2 + \ii\omega\tau^{-1}) / \mu_c \,,
\end{align}
where $D = \pi\epsilon_0\hbar^2/e^2$, and the substitution $\sigma \approx \sigma^{\rm intra}$ has been made in the last step.~\cite{jablanprb80} Separating Eq.~\eqref{betaapprox} into real and imaginary parts, we find $\beta' = D (\epsilon_1 + \epsilon_2)\omega^2 / \mu_c$, and $\beta'' = D (\epsilon_1 + \epsilon_2)\omega\tau^{-1} / \mu_c$. 
This gives us some insight to the physical characteristics of the A mode. 
First, similar to the case of SPs supported on metal films, the A mode is better confined (larger $\beta'$) for higher values of $\epsilon_1$ or $\epsilon_2$ (as seen in Fig.~1(b)), accompanied with a proportional increase in absorption loss (larger $\beta''$). 
Second, $\beta'$ varies as $\lambda^{-2}$, decreasing at a rate much faster than $\beta''$ which varies as $\lambda^{-1}$. Therefore, where the approximation is valid, $n'$ varies as $\lambda^{-1}$ and $n''$ is constant, in agreement with the dispersion curves of Fig.~1(b).
Third, a higher doping level $\mu_c$ decreases the confinement, which can be understood because the graphene becomes more like a perfect conductor with increased conductivity (see Eq.~\eqref{sigintrab}) and supports SP modes whose fields extends more into the dielectric media. 
This trend is shown in Fig.~2(a) with the case of isolated monolayer graphene ($\epsilon_1 = \epsilon_2 =1$), where $n'$ for different values of $\mu_c$ is shown. It can be seen that the approximate solution Eq.~\eqref{betaapprox} is in excellent agreement with 
the exact dispersion relation~\eqref{spdispigi} except near the plasmon resonance frequency. 
For antisymmetric A modes well-described by the approximation~\eqref{betaapprox}, the normalized propagation length $L/\lambda_{sp}$ at which the 
field amplitude of the SP falls to $1/e$ of the initial value is~\cite{jablanprb80}
\eq\label{fom}
L/\lambda_{sp} \approx \omega\tau/2\pi = c\tau/\lambda\,,
\eeq
which depends only on the operating wavelength and relaxation time $\tau$. As seen from the inset of Fig.~2(b), the approximate form 
$c\tau/\lambda$ remains highly accurate for operating wavelengths up to $\sim 30 \micron$. 
The slight deviation for $\lambda \gtrsim 30 \micron$ is attributed to the inequality $\beta'' << \beta'$ in the approximation~\eqref{betaapprox}, as the inequality is 
less accurate for longer wavelengths  since as explained, $\beta' \propto \lambda^{-2}$ and $\beta'' \propto \lambda^{-1}$.
The peak value of $L/\lambda_{sp}$ is seen to increase with $\mu_c$ according to the curve $c\tau/\lambda$ because a higher chemical 
potential has the effect of blue-shifting $\lambda_p^{\rm (eff)}$, see Eq.~\eqref{effplasmonfreq}. 
It is worthwhile noting that the ratio $L/\lambda_{sp} = n'/2\pi n''$ can also act as a figure-of-merit (FOM) for 1D  plasmonic waveguides,~\cite{beriniopex} and is shown in Fig.~2(c) for $0.2 \eV \leq \mu_c \leq 1.2 \eV$ as a function of wavelength. However, in the case of graphene, $L/\lambda_{sp}$ is virtually independent of important parameters like $\mu_c, \epsilon_1,$ and $\epsilon_2$ (see Eq.~\eqref{fom}). As such we propose an alternative FOM,
\eq\label{newfom}
{\rm FOM} = L/\sqrt{\lambda\lambda_{sp}} = \sqrt{n'}/2\pi n'' \,,
\eeq 
where the propagation length of the SP is normalized by the geometric mean of its effective wavelength and the free-space wavelength. 
The FOM as defined in Eq.~\eqref{newfom} for $0.2 \eV \leq \mu_c \leq 1.2 \eV$ as a function of wavelength is shown in Fig.~2(d). In contrast to the ratio $L/\lambda_{sp}$, the FOM is clearly distinct for different values of the chemical potential and wavelength. As such, we propose the newly defined FOM as a measure of the performance of graphene plasmon waveguides. For the graphene pair studied in the next section, this FOM will also be applied to quantify its performance as a SP waveguide. 

%Here $\delta = \lambda/[2\pi\Re(\alpha)]$ is the transverse distance from the graphene surface at which the SP field has decay to $1/e$ of the peak value. 

\section{Tuning the SP mode with a pair of parallel graphene layers}
Next, we demonstrate how one may tune the confinement and propagation characteristics of the SP mode by introducing an additional layer of graphene. Here, two monolayers of graphene (at $x = 0,\, d$) are separated by a 
thin dielectric medium ($\epsilon_d$) as shown in Fig.~3(a). 
Let us neglect the effects of 
absorption losses due to coupling with optical phonons 
($\lambda^{\rm op} \approx 6.3~\micron$) as described in Sect. I and consider the range of wavelengths $\lambda^{\rm op} \leq \lambda \leq 50~\micron$.
We are primarily concerned with graphene pairs whose spatial separation $d$ is sufficiently large such that effects of interlayer hopping, which can be important for closely-spaced graphene pairs (few angstroms apart) or bilayer graphene, have no significant bearing on the SP dispersion.~\cite{sarmaprb82} For the following analysis, the smallest separation $d$ we 
consider is $d_{\rm min}/\lambda = 0.001$, i.e. $d_{\rm min} \approx 6 \nm$. 

For a parallel pair of graphene monolayers separated by a distance $d$, the tangential $H$-field in each of the dielectric media can be expressed as
\begin{equation}\label{hfieldtmpp}
H = \left \{ \begin{array}{ll}
N_1 H_{1} \,\, \exp(\alpha_1 x + \ii \beta z) \, & \textrm {$x < 0$ } \,,\\
N_1 (H_a \cosh\alpha_dx + H_b \sinh\alpha_dx ) \exp(\ii \beta z) \, & \textrm {$0 \leq x \leq d$ } \,,\\
N_1 H_{2} \,\, \exp(-\alpha_2 (x-d) + \ii \beta z) \, & \textrm {$x > d$ } \,,
\end{array} \right.
\end{equation} 
where $N_1$ is a normalization constant similar to $N_0$ in Eq.~\eqref{hfieldtm1}. 
Matching the boundary conditions at $x = 0$ and $x = d$, we obtain for the tangential $H$-field
\begin{subequations}\label{hgppbc}
\begin{align}
H_a & = (1+ \frac{\ii \sigma_1 \alpha_1}{\omega \epsilon_1 \epsilon_0})H_1\,, \quad (x = 0) \label{hgppbcx0} \\
H_a \cosh\alpha_dd + H_b \sinh\alpha_dd & = (1+ \frac{\ii \sigma_2 \alpha_2}{\omega \epsilon_2 \epsilon_0})H_2\,, \quad (x = d) \label{hgppbcxd}
\end{align}
\end{subequations} 
and for the tangential $E_z$ component,
\begin{subequations}\label{ezppbc}
\begin{align}
\frac{\alpha_d}{\epsilon_d \epsilon_0} H_b & = \frac{\alpha_1}{\epsilon_1 \epsilon_0} H_1\,, \quad\,\,\,\,\, (x = 0) \label{ezppbcx0} \\
\frac{\alpha_d}{\epsilon_d \epsilon_0} (H_a \sinh\alpha_dd + H_b \cosh\alpha_dd) & = -\frac{\alpha_2}{\epsilon_2 \epsilon_0} H_2\,, \quad (x = d) \label{ezppbcxd}
\end{align}
\end{subequations} 
where $\alpha_d = \beta^2-\epsilon_d k_0^2$. In Eq.~\eqref{hgppbc}, $\sigma_1$ and $\sigma_2$ are the conductivities of the graphene monolayers at $x = 0$ and $x = d$. The graphene conductivities $\sigma_1$ and $\sigma_2$, which are calculated with Eq.~\eqref{kubo}, may be tuned through their respective chemical potentials $\mu_1$ and $\mu_2$. Eliminating $H_1, H_2, H_a,$ and $H_b$ from Eqs.~\eqref{hgppbc}~and~\eqref{ezppbc} yields the dispersion relation
\eq\label{ppdisp}
\tanh\alpha_dd = -\frac{\Gamma_1 + \Gamma_2}{1 + \Gamma_1 \Gamma_2} \,,
\eeq
with 
\eq\label{biggam}
\Gamma_j = \frac{\alpha_d\epsilon_j}{\alpha_j\epsilon_d} (1 + \frac{\ii\sigma_j\alpha_j}{\omega\epsilon_j\epsilon_0}) \,.
\eeq
In the symmetric case where $\epsilon_1 = \epsilon_2$ and $\sigma_1 = \sigma_2 \,(\mu_1 = \mu_2)$, the dispersion relation~\eqref{ppdisp} splits into two branches
\begin{subequations}\label{ppdisp2br}
\begin{align}
\tanh\frac{\alpha_dd}{2} & = - \frac{1}{\Gamma_1}\,, \quad \textrm {(even, AS)} \label{ppdispeve} \\
\coth\frac{\alpha_dd}{2} & = - \frac{1}{\Gamma_1}\,, \quad \textrm {(odd, AA)} \label{ppdispodd}
\end{align}
\end{subequations} 
where $\Gamma_1 = \Gamma_2$ due to symmetry. 
Let us take the case $\epsilon_1 = \epsilon_2 = \epsilon_d = 1$, which is a configuration that has also been studied in the context of Casimir forces~\cite{woodsprb82} and quasi-transverse electromagnetic waveguide modes~\cite{hansonjap104}. 
Unless otherwise specified, it will be taken that the geometry is fully symmetric with $\mu_1 = \mu_2 = \mu_c$, and $\epsilon_1 = \epsilon_2 = \epsilon_d = 1$. 

For large separations $d$ of the graphene pair, both Eqs.~\eqref{ppdisp}~and~\eqref{ppdisp2br} reduce to the dispersion relation~\eqref{spdispigi} for 
the case of the I-G-I. 
For sufficiently small $d$ such that the SP supported on individual graphene monolayer interacts with each other, the dispersion splits into 
an even and an odd mode.
The typical field distribution of the $H$-field for these two modes  
are shown for $d = 0.01\lambda$, $\lambda = 10 \micron$, and $\mu_c = 0.4 \eV$ as an example in Fig.~3(b). The $H-$field displays a symmetric (antisymmetric) character across the gap for the even (odd) mode. Consistent with the analysis for monolayer graphene in Sect. II, the $H-$field is antisymmetric across each graphene layer. As such, let us label the even mode AS and the odd mode AA, where the first A in the labels refers to the antisymmetric $H$-field distribution across each graphene monolayer.
Additionally, let us note that the other two branches SA and SS (the first label S refers to the symmetric $H$-field across the graphene), can also be supported when the dielectric layer $(\epsilon_d)$ is a thin metal film.~\cite{tamir} This case is verified with numerical simulations (not shown), where it was found that, similar to the case of the I-G-I, the dispersion with and without the graphene films differ only marginally. As such, we focus on the I-G-I-G-I structure.

Figure~3(c) illustrates the splitting of the SP dispersion into the AA (thin blue curves) and AS (thick red curves) modes as 
the separation $d$ is decreased, for the case $\lambda = 10 \micron$ and $\mu_c = 0.4 \eV$.
Notice the difference in the vertical scales for $n'$ and $L/\lambda_{sp}$.
For sufficiently large $d$, both the AA and AS branches merge to the A mode supported by the monolayer graphene.
As the separation $d$ is decreased, the real part of the effective index $n'$ increases (decreases) monotonically for the AS (AA) branch, corresponding to improved (degraded) confinement of the SP mode. 
On the other hand, the imaginary part of the effective index $n''$ (inset of Fig.~3(c)) for both modes exhibits a nonmonotonic behavior. While this nonmonotonic behavior is also observed in the normalized propagation length $L/\lambda_{sp}$ for the AA branch (dashed blue curve), it is absent for the AS branch (dashed red curve) however. This is because the SP wavelength decreases more quickly than the absorption loss increases as $d$ is decreased. 
At very small separations $d$, or equivalently long wavelengths $\lambda$, the AS branch would be strongly suppressed due to the exponential increase in absorption.

For the AA mode, it is expected, within linear optics, that the amplitude of the $H$-field across the dielectric gap ($\epsilon_d$) approaches asymptotically to a saw-tooth waveform with a near-zero amplitude as the separation $d$ decreases (see Fig.~3(b)). This implies that the electromagnetic field in the gap is squeezed out of the gap into the two dielectric half-space as $d$ decreases, explaining the decrease in the confinement ($n'$) in Fig.~3(c). 
In the limit $d \longrightarrow 0$, the AA mode tends to the A mode for the case of the bilayer graphene.~\cite{fnote1} Here we estimate the conductivity of the bilayer graphene to be twice the conductivity of the monolayer, following the approximation $N\sigma$ for few layer graphene,~\cite{ferrarinphot,hansonjap104,ferrarinl} where $N$ is the number of layers $(N < 6)$. In Fig.~3(c), it is seen that $n'$ for the AA branch approaches that of the bilayer case, marked with a circle on the vertical axis. For $\mu_c = 0.4 \eV$ and $\lambda = 10 \micron$, the SP dispersion of the monolayer graphene satisfies the condition $\beta'' << \beta'$ (see Fig.~2(a)), and thus $L/\lambda_{sp}$ for the bilayer case and monolayer case are almost equal (see inset Fig.~2(b)). These observations are all in agreement with the data for the AA mode represented by the blue curves in Fig.~3(c).  
For the AS mode, as long as the separation $d$ is not too small ($d \gtrsim 0.01\lambda$ for this example), the increase in the confinement is accompanied with reduced absorption loss ($n''$). This concomitant improvement in confinement and absorption loss, which is contrary to the trade-off between confinement and loss associated with conventional SP modes,~\cite{bozhenphot} is also observed for SPs supported in dielectric-loaded metallic waveguides~\cite{bozheprb75} that exhibit similar  dispersion characteristics as the AS mode.

The influence of the operating wavelength ($\lambda$) and chemical potential ($\mu_c$) on the dispersion of the AA and AS modes are shown in Figs.~4 and 5, respectively. 
As observed, the behavior of the AS and AA branches share a number of similar characteristics with the case of the monolayer graphene investigated above. 
Let us first focus on the influence of the wavelength $\lambda$ (Fig.~4), where the chemical potential is taken to be $\mu_c = 0.4 \eV$. 
For both branches, Figs.~4(a) and 4(b) show that $n'$ is inversely proportional to $\lambda$, whereas $n''$ changes only weakly for different wavelengths (Figs.~4(c) and 4(d)). The result is an overall decrease in the normalized propagation length $L/\lambda_{sp}$ as $\lambda$ increases. 
Additionally, because an increased wavelength effectively decreases the separation $d$, the two branches begin to merge for larger values of the normalized separation $d/\lambda$. 

Figure 5 shows the influence of the chemical potential $\mu_c$ on the dispersion of the two modes, where the wavelength $\lambda$ is taken to be $10 \micron$. 
Both $n'$ (Figs.~5(a) and 5(b)) and $n''$ (Figs.~5(c) and 5(d)) decreases as $\mu_c$ increases, i.e., a higher chemical potential increases the conductivity of graphene, thereby reducing the confinement and the absorption loss. 
For a higher value of the chemical potential, because of the longer 
effective SP wavelength ($\lambda_{sp} = \lambda/n'$), the separation between the 
graphene pair is effectively decreased, and thus the two branches start merging at larger separations $d/\lambda$. 
A common feature of all the dispersion curves in Figs.~4 and 5 is that the change in $n$ or  $n''$ becomes increasingly less profound as $\lambda$ or $\mu_c$ increases. 
These behaviors suggests an inversely proportional relationship between the wavenumber $\beta$ of each SP mode (AA or AS) and $\lambda$ or $\mu_c$, in a 
similar fashion to Eq.~\eqref{betaapprox} for the monolayer case in Sect. II.
From the above observations, the influence of $\epsilon_d$ on the dispersion curves of the two modes may also be inferred. Since increasing $\epsilon_d$ has the same effect of reducing the effective SP wavelength $\lambda_{sp}$, the behavior for a greater dielectric constant of the separating medium ($\epsilon_d > 1$) is qualitatively similar to a lower value of the chemical potential. As such, the AS and AA branches would merge for smaller separation $d/\lambda$ for a higher value of $\epsilon_d$ (with all other parameters kept constant). Similar to a lower chemical potential, a higher $\epsilon_d$ would result in a higher maximum value of $n''$. 
These behaviors for different values of $\epsilon_d$ have been verified with calculations (not shown) .

Next, let us compute the FOM as defined in Eq.~\eqref{newfom} for the AA and AS modes.
We shall make a comparison between the FOM and the ratio $L/\lambda_{sp}$ to see if the newly-defined FOM serves better to quantify the performance of the two modes for waveguiding operation. 
The two quantities $L/\lambda_{sp}$ and FOM for the two modes as a function of the normalized separation $d/\lambda$ and 
wavelength $\lambda$ are shown in Fig.~6, with the chemical potential taken to be $\mu_c = 0.8 \eV$. 
Except for a narrow region confined to $d/\lambda \lesssim 0.1$ and $\lambda \lesssim 20 \micron$, Fig.~6 (top panels) reveals that 
the ratio $L/\lambda_{sp}$ varies only very weakly with $d/\lambda$ for a given wavelength. 
Furthermore, the ratio $L/\lambda_{sp}$ fails to highlight the fact that the AS mode is extremely lossy 
for small separations $d$. By looking at the top panels of Fig.~6, one might also be misled to believe that the AA and AS modes are not useful for waveguiding operation for the long wavelengths in the terahertz regime. These deficiencies are circumvented with the corresponding FOM shown in the bottom panels of Fig.~6. 
First, let us note that unlike the ratio $L/\lambda_{sp}$, the FOM of the two modes for a given set of parameters lie in a similar range of values (see the scale on the colorbars). This is due to the factor $\sqrt{n'}$ in the FOM of Eq.~\eqref{newfom} instead of $n'$ in the ratio $L/\lambda_{sp}$.
Second, the FOM of the AA mode for very small separations $d$, or equivalently long wavelengths, can be relatively high. This is due primarily to the smaller absorption loss as the AA mode tends to the bilayer case. On the other hand, the AS mode provides a better FOM for 
intermediate values $0.05 \lesssim d/\lambda \lesssim 0.2$ and $15 \micron \lesssim \lambda \lesssim 45 \micron$.
Thus for a properly chosen separation $d$, the parallel graphene pair offers at least one plasmon mode that is suitable for waveguide operation over the mid-infrared and terahertz spectrum.
For a higher (lower) $\mu_c$, the absorption loss $n''$ decreases (increases), confinement degrades (improves), and the general trend is an increase (decrease) in the FOM with the two branches starting to merge at larger (smaller) separations $d/\lambda$.

So far we have considered fully symmetric geometries. 
Yet another degree of freedom to tune the SP dispersion of the two branches is to 
control the chemical potential of one of the graphene monolayer while 
keeping the other one fixed. Evidently when $\mu_1 \neq \mu_2$, the symmetry is  broken, 
and the fields of the modes no longer obey even or odd symmetry (see Figs.~7(a) and 7(b)). 
As such, let us define new branches AS' and AA' that stem from the AS and AA modes (respectively) for the symmetric case where $\mu_1 = \mu_2$. 
As an example, let us take that $\mu_1 = 0.4 \eV$ is fixed with $\mu_2$ adjustable for an operating wavelength $\lambda = 30 \micron$. 
For this case, the two modes for which $\mu_1 = \mu_2 = 0.4 \eV$ are the AA and AS modes, whose dispersion has been given in Fig.~4 (red dotted curves). 
The influence on the dispersion when $\mu_2$ is varied so that $\mu_1$ and $\mu_2$ are no longer equal is shown in Figs.~7(c) to 7(f). 
To aid the analysis, let us further categorized the dispersion behavior of the AS' and AA' branches for $\mu_2 < \mu_1$ and $\mu_2 > \mu_1$.
For the AS' (AA') branch, the dispersion is only weakly influenced for $\mu_2 > \mu_1$ ($\mu_2 < \mu_1$), but changes visibly for $\mu_2 < \mu_1$ ($\mu_2 > \mu_1$). 
Figures~7(c) and 7(d) show that $n'$ generally decreases for increasing $\mu_2$ for both branches, consistent with earlier analysis on the AA and AS modes (see Figs.~4 and 5). 
As observed in Figs.~7(e) and 7(f), the dispersion of $n''$ for the AA' and AS' branches exhibits a non-monotonic 
behavior, also similar to their symmetric counterparts. 
For the AS' branch with $\mu_2 < \mu_1$ and the AA' branch with $\mu_2 > \mu_1$, both $n'$ and $n''$ decrease for increasing $\mu_2$ as expected. 

The dispersion behavior is however a little different for the weakly influenced case.  
In this case, for a qualitative explanation of the dispersion, let us take the AS' (AA') branch with $\mu_2 > \mu_1$ ($\mu_2 < \mu_1$) to be a mode that has characteristics dominant of the AS (AA) mode but with a slight perturbation due to contribution from the AA (AS) mode, i.e., the two modes are mixed but with a dominant AS or AA character. 
For the AS' branch, $n''$ is decreased very slightly for increasing $\mu_2$ for $d/\lambda \lesssim 0.02$.
In the range $0.02 \lesssim d/\lambda \lesssim 0.12,$ the AS' mode is 
modified with contributions from the more lossy AA mode, thus the absorption loss $n''$ is increased even for $\mu_2 > \mu_1$.
For the AA' branch, $n''$ is increased slightly for decreasing $\mu_2$ for $d/\lambda \lesssim 0.04$. 
In the range $0.04 \lesssim d/\lambda \lesssim 0.12,$ the AA' mode is 
modified with contributions from the less lossy AS mode, thus decreasing the absorption loss $n''$ for $\mu_2 < \mu_1$.

We have shown the typical mode profile for the AS' and AA' modes for $\mu_2 \gtrless \mu_1$ in Figs.~7(a) and 7(b).  
In agreement with the analogy of the mixed mode, it is seen that the case of the weakly 
influenced dispersion ($\mu_2 = 0.8 \eV$ for AS' given in dashed blue curve of Fig.~7(a) and , $\mu_2 = 0.3 \eV$ for AA' given in solid red curve of Fig.~7(b)) displays a 
relatively stronger AS or AA behavior. On the other hand, the solid red curve in Fig.~7(a) ($\mu_2 = 0.2 \eV$) resembles that of the A mode (for monolayer graphene) with very little influence from the other graphene layer, and the dashed blue curve in Fig.~7(b) ($\mu_2 = 0.8 \eV$) also resembles that of the A mode but {\em slightly split} within the graphene pair. 
A complete understanding of the physical mechanism leading to the weakly dispersive behavior of the AS' and AA' modes for the somewhat contrasting conditions described above is a future topic of investigation. 
The above analysis also shows that, while an asymmetric chemical potential generally increases the absorption loss for the AS' branch, it may serve to reduce the absorption loss for the AA' branch quite significantly. 
The corresponding FOM for the two branches are shown in Figs.~7(g) and 7(h), where it is seen that the dispersion characteristics described above are well-captured with the FOM. While the highest FOM is associated with the AA' branch for large values of $\mu_2$ ($\mu_2 > \mu_1$), the lowest FOM is observed for the AS' branch for small values of $\mu_2$ ($\mu_2 < \mu_1$).

\section{Conclusion}
In summary, for this article we have revisited the plasmon disperison in monolayer graphene, followed by 
a thorough investigation of the dispersion of the SP modes supported in a parallel pair of graphene monolayers separated by a distance $d$. 
Due to the matching of boundary conditions across the interface of the monolayer graphene, it is found that the I-G-I combination supports only the A mode (antisymmetric magnetic field across the monolayer).
To quantify the waveguiding performance of the SP mode, we found that a newly-defined FOM $=L/\sqrt{\lambda\lambda_{sp}} = \sqrt{n'}/2\pi n''$ is more optimal than the ratio  $L/\lambda_{sp}$ because for a wide spectral range for which the approximation~\eqref{betaapprox} is satisfied, the latter is almost independent of important parameters such as the chemical potential and dielectric constants of the neighboring media.
 
In the case of the graphene parallel pair, the separation $d$ is employed as an additional degree of freedom to tune the SP dispersion. 
For the fully symmetric geometry, the dispersion for the graphene pair splits into an even AS and odd AA branch for sufficiently small separations $d$. 
While the AS and AA modes may be characterized based on the symmetry of each mode, they share 
similar characteristics as the A mode supported in monolayer graphene. 
For instance, the influence of the wavelength and the chemical potential on the dispersion of the 
AS and AA modes are similar to that 
of the A mode in monolayer graphene, i.e. their wavenumbers are somewhat inversely proportional to the wavelength and chemical potential, similar to Eq~\eqref{betaapprox}.
The non-monotonic behavior of the absorption loss ($n''$) means that there exist a 
certain range whereby the AS and AA modes are less absorptive compared to the monolayer case. 
By calculating the FOM, it is seen that the AA mode is optimal for small separations $d$ or equivalently, long wavelengths $\lambda$. 
On the other hand, the AS mode offers a better FOM than the AA mode for an intermediate range of separations and wavelengths where it is both less absorptive and better confined. 
For very small $d$ however, the AS mode is strongly suppressed as its absorption increases exponentially. 
It is noted that the AA and AS branches here correspond to the optical and acoustic plasmon branches (respectively) in 
existing literature on theoretical investigation of the 
plasmon dispersion relation of the parallel graphene pair.~\cite{sarmaprb75, sarmaprb80, sarmaprb82}
Varying the chemical potential of one of the graphene monolayers allows for yet another degree of freedom to tune the SP dispersion. 
Strictly speaking, the fields no longer possess symmetry in this case. However, we 
take the case $\mu_1 = \mu_2$ as reference, and consider branches AS' and AA' that stem from the AS and AA modes respectively.
Our analysis shows that, under the right conditions, an asymmetric chemical potential can significantly reduce the absorption loss for the AA' branch. 
Our results demonstrate the potential to effectively tune the SP mode dispersion with a parallel pair of graphene monolayers to achieve well-confined and propagative SP modes in the mid-infrared and terahertz regime with graphene structures. 
This graphene plasmon waveguide platform is promising for mid-infrared and terahertz applications.

\subsection*{Acknowledgements} This work was supported by the Agency for Science and Technology Research (A*STAR), Singapore, Metamaterials-Nanoplasmonics research program under A*STAR-SERC grant No. 0921540098.

\newpage

\begin{figure}[ht]
\centering
\epsfig{file=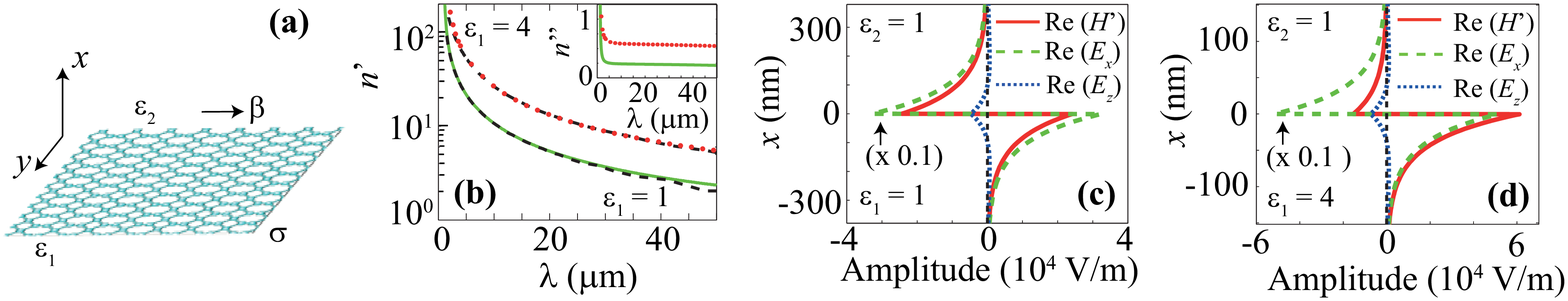, width=5.0 in}
\caption{SP dispersion characteristics for I-G-I structure, with $\epsilon_2 = 1$. The geometry is illustrated in {\bf (a)}. The real part of the SP effective index $n'$ for $\epsilon_1 = 1$ (green solid curve) and $\epsilon_1 = 4$ (red dotted curve) as a function of wavelength is shown in {\bf (b)}, inset: the corresponding imaginary part $n''$. In {\bf (b)}, the colored curves are results from Eq.~\eqref{spdispigi}, and black dashed curves depict calculations of the SP dispersion relation with the RPA~\cite{sarmaprb75} shown here for comparison. Field components ($E_x, E_z, H' = \eta_0 H$) with $\lambda$ taken to be $8 \micron$ are 
shown for {\bf (c)} $\epsilon_1 = 1$ and {\bf (d)} $\epsilon_1 = 4$. In {\bf (c)} and {\bf (d)}, the normal electric field component $E_x$ is multiplied by a factor of 0.1 for the purpose of illustration. The chemical potential $\mu_c$ is taken to be $0.8 \eV$ for all calculations. }
\end{figure}

\begin{figure}[ht]
\centering
\epsfig{file=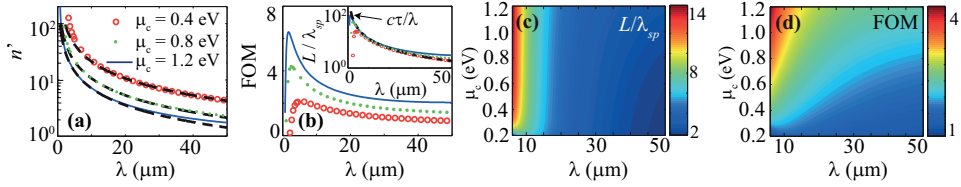, width=5.0 in}
\caption{SP dispersion characteristics for isolated monolayer graphene ($\epsilon_1 = \epsilon_2 = 1$). {\bf (a)} $n'$ and {\bf (b)} FOM based on Eq.~\eqref{newfom} for several values of the chemical potential. The legend in {\bf (a)}  applies to {\bf (b)}. In {\bf (a)}, black dashed curves depict the approximate solution in Eq.~\eqref{betaapprox}. Inset of {\bf (b)}: normalized propagation length $L/\lambda_{sp}$ and the approximation $c\tau/\lambda$ (black dashed curve). {\bf (c)} $L/\lambda_{sp}$ and {\bf (d)} FOM as a function of the chemical potential and wavelength.}
\end{figure}

\begin{figure}[ht]
\centering
\epsfig{file=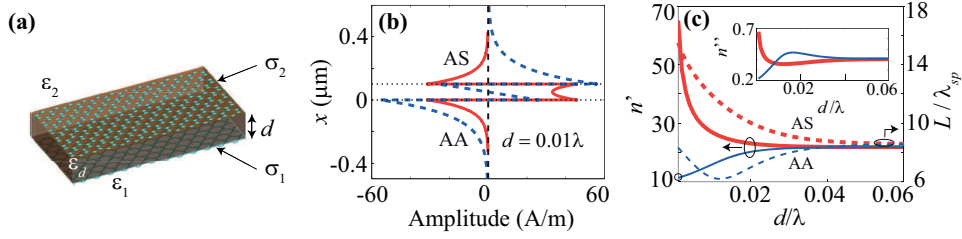, width=5.0 in}
\caption{Splitting of the SP dispersion into the AA and AS modes with 
the geometrically symmetric I-G-I-G-I structure ($\epsilon_1 = \epsilon_2 = \epsilon_d = 1, \mu_1 = \mu_2 = \mu_c$), illustrated here 
for the case $\lambda = 10 \micron$, and $\mu_c = 0.4 \eV$. {\bf (a)} Depicting the geometry. {\bf (b)} Plot of the typical magnetic field component of the symmetric AS and antisymmetric AA mode 
with the separation $d$ taken to be $0.01\lambda$. {\bf (c)} Dispersion curves as a function of the normalized 
separation $d/\lambda$ for the AA (thick red curves) and AS (thin blue curves) mode. The circle on the left vertical axis in {\bf (c)} indicates $n'$ for the bilayer (I-2G-I) case. }
\end{figure}

\begin{figure}[ht]
\centering
\epsfig{file=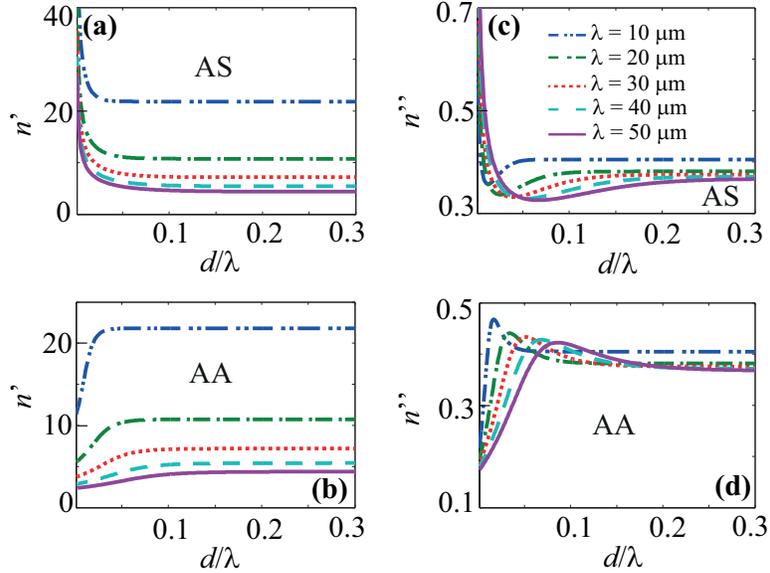, width=4.0 in}
\caption{Influence of the operating wavelength on the dispersion of the AA and AS modes. Dispersion curves for $n'$ ({\bf (a)} and {\bf (b)}) and $n''$ ({\bf (c)} and {\bf (d)}) of the AS and AA modes are shown in the top and bottom panels respectively, with the chemical potential $\mu_c$ taken to be $0.4 \eV$. The legend in {\bf (c)} applies to all other subfigures.}
\end{figure}

\begin{figure}[ht]
\centering
\epsfig{file=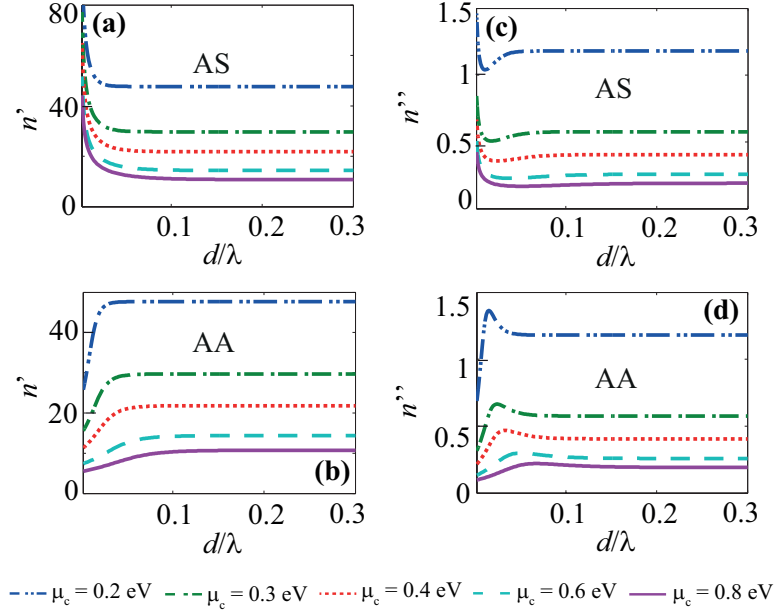, width=4.0 in}
\caption{Influence of the chemical potential on the dispersion of the AA and AS modes. Dispersion curves for $n'$ ({\bf (a)} and {\bf (b)}) and $n''$ ({\bf (c)} and {\bf (d)}) of the AS and AA modes are shown in the top and bottom panels respectively, with the wavelength $\lambda$ taken to be $10 \micron$. The legend is shown at the bottom of the figure.}
\end{figure}

\begin{figure}[ht]
\centering
\epsfig{file=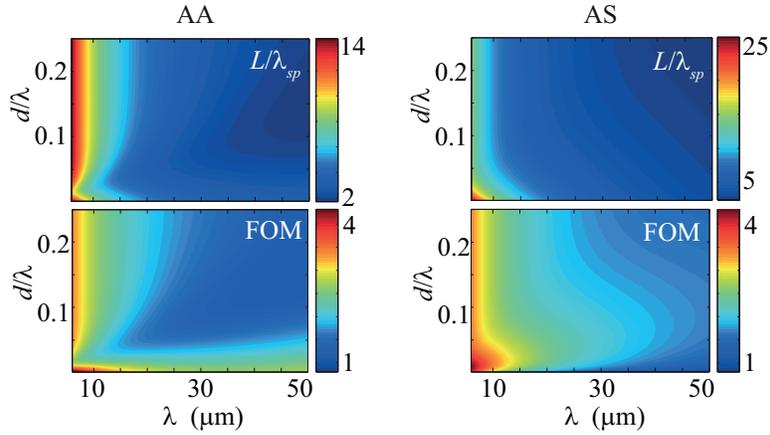, width=4.0 in}
\caption{\label{f_pcqbi}Waveguiding performance of the AA and AS modes as a function of wavelength $\lambda$ and normalized separation $d/\lambda$, with $\mu_1 = \mu_2$ taken to be $0.8 \eV$. Top panels: $L/\lambda_{sp}$, bottom panels: FOM. }
\end{figure}

\begin{figure}[ht]
\centering
\epsfig{file=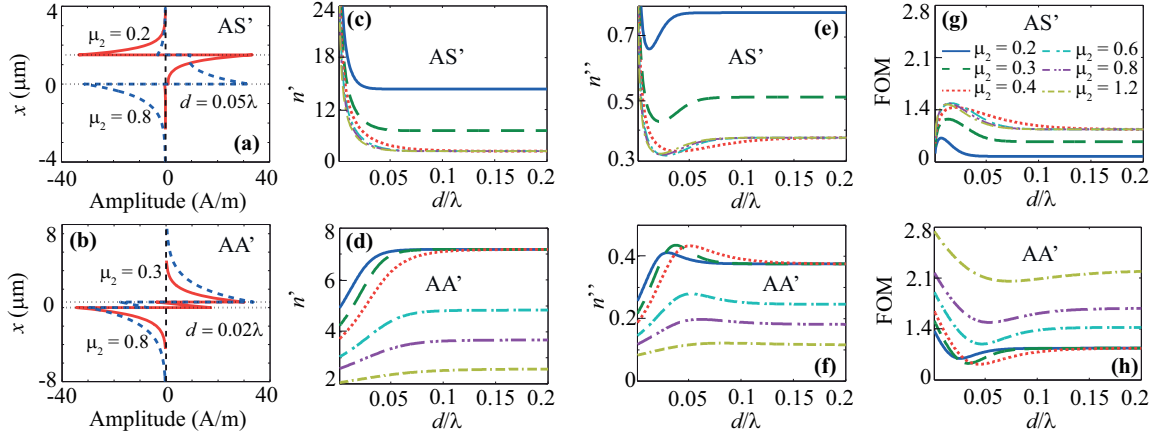, width=6.0 in}
\caption{\label{f_pcqbipp}Influence on the SP dispersion when the chemical potential on one of the graphene sheets is varied ($\mu_1 \neq \mu_2$). It is taken that $\mu_1 = 0.4 \eV$ is fixed, while $\mu_2$ varies from $0.2 \eV$ to $1.2 \eV$. Strictly speaking, the fields no longer possess symmetry for $\mu_1 \neq \mu_2$, as can be seen in {\bf (a)} and {\bf (b)}. However, we 
take the case $\mu_1 = \mu_2 = 0.4 \eV$ as reference, and consider branches AS' and AA', which stem from the AS and AA modes respectively. 
Dispersion curves for $n'$ ({\bf (c)} and {\bf (d)}), $n''$ ({\bf (e)} and {\bf (f)}), and the corresponding FOM of Eq.~\eqref{newfom} ({\bf (g)} and {\bf (h)}) as a function of $d/\lambda$ for various values of $\mu_2$ for AS' and AA' are given in the top and bottom panels, respectively. The legend in {\bf (g)}, where the chemical potential $\mu_2$ is specified in units of eV, applies to subfigures {\bf (c)} - {\bf (h)}. The wavelength is taken to be $\lambda = 30 \micron$. }
\end{figure}

\end{document}